\documentclass[%
 reprint,
 amsmath,amssymb,
 aps,
pra,
]{revtex4-1}

\usepackage{graphicx}% Include figure files
\usepackage{dcolumn}% Align table columns on decimal point
\usepackage{bm}% bold math

\usepackage{diagbox}
\usepackage{multirow}
\usepackage{hyperref}

\usepackage{CJKutf8}

\begin{document}

%\begin{CJK*}{utf8}{}
\begin{CJK}{UTF8}{gbsn}

\title{Non-Hermitian curved space via inverted wave equation}

\author{{C. Zhang (张春娟)}\textsuperscript{1}}
\author{{M. Ikram (李克龙)}\textsuperscript{1}}
\author{{Y. Liu (刘泱杰)}\textsuperscript{1}}
\email[Corresponding author: ]{yangjie@hubu.edu.cn}
\author{{H. Lin}\textsuperscript{2}}
\author{{Bin Zhou}\textsuperscript{1, 3}}

\affiliation{
$^{1}$ Department of Physics, School of Physics, Hubei University, Wuhan 430062, P. R. China\\
$^{2}$ College of Physical Science and Technology, Central China Normal University, Wuhan 430079, P. R. China\\
$^{3}$ Wuhan Institute of Quantum Technology, Wuhan 430206, P. R. China}%

%\date{v28, 15th Feb. '26 submitted to \emph{Opt. Lett.}, manuscript ID \#591962. }
%submitted to PNFA 26th Apr., major revision decision received 27th Jun., revised 7th Aug., \\
%minor amendment received 10th Sept, revised 18th and 20th Sept. '26, 
%\today
\date{resubmitted 20th Sept. '26}

\begin{abstract}

	Directly determining graded materials from prescribed wave amplitude and phase is an inverse-design method developed following transformation optics (TO), which enables reflectionless propagation for an incidence wave. However, exiting formulations of this inverted method yields Hermitian media and are thus not applicable to non-Hermitian (NH) photonics. In this Letter, we extend this method to NH media, thereby offering more freedom for manipulating waves of no reflection. Our picture of curved-space powered with spatially distributed gain and loss, is exemplified by three types: amplitude control, phase conversion, and direction shunting. These examples showcase precise wave manipulation in a surprisingly simple manner, which is implementable in photonic platform using similar techniques employed in TO.

\end{abstract}

%\keywords{Wave equation, metamaterials, reflectionlessness.}%Use show keys class option if keyword
                              %display desired
\maketitle

\end{CJK}

%\tableofcontents

\section{Introduction}

To precisely control waves in both amplitude and phase, one inversely designs inhomogeneous metamaterials based on the known wave forms, which saves the computation demands forwardly performed along with trial-and-error. Influenced by transformation optics (TO)~\cite{Pendry2006Controlling, Ulf2010Geometry}, we directly solve material profiles by predefining waves without mapping from virtual space, which offered a new strategy of molding electromagnetic waves using isotropic materials~\cite{Philbin2014Making, Vial2016class, Philbin2016All-freq, Y2017Direct, King2018scattering-free, YuS2018Bohmian, Makris2020Scattering-free, Kresic2021Confinement, Biswas2024Reflectionless}. This inverse-design theory not only bypasses the requirement for mother design before mapping, but also provides further manipulation on phase control~\cite{Y2017Direct, King2018scattering-free}, which were usually not aimed for by other methodologies. However, this direct method to invert the wave equation relies on the nontrivial techniques of solving partial differential equations (PDE) from separation of amplitude and phase~\cite{Vial2016class, Y2017Direct}. Moreover, the resulting material parameters were typically taken passive, with no source or sink. This assumption no longer applies in the non-Hermitian (NH) scenarios~\cite{El2018NH_phys, Horsley2015Spatial, Steinfurth2022ph_constant-int} with both loss and gain to reshape the wave otherwise. This motivates us to encode the non-Hermiticity into the inverse design framework and thus gives rise asymmetric control of wave flow, for example an asymmetric transmitter which switches on in discrete directions and off otherwise as an essential piece for optical switch devices.

Surprisingly, an isotropic material profile embracing loss and gain can be simpler to design than its Hermitian counterpart, because it sidesteps the mathematical difficulty of separation of amplitude and phase~\cite{Philbin2014Making, Vial2016class, King2018scattering-free}. Inspired by the synergy between NH photonics and TO~\cite{Kresic2022Transforming}, we extend the material parameters to a \emph{complex} function of space to embrace distributive loss and gain, and then develop a curved-space analogue picture in the spirit of TO for predefined incidence. The merit of such extension to complex value is two-fold: (1) it encapsulates spatial gain and loss within access to metamaterial engineering~\cite{Ye2017reflectionless}, in which one uses local metamaterial elements to realise the discretized refractive index profiles; and (2) the complex-valued material profile naturally hosts complex-valued waves of no reflection, which shall immediately clarify the inverting method as detailed in Eq.~\eqref{eq:central} of Sec. II below. In Sec. III, our three types of examples are numerically verified to work per design purposes: amplitude control, phase conversion, and direction shunting. It is noted that our theory applies to other types of wave forms such as under paraxial approximation and for higher dimension, and puts no explicit restraint on boundary conditions. Finally in Sec. IV, we conclude our Letter by envisioning its universality for controlling non-reflection waves in the NH landscape. By implementing absorbing and lasing techniques established in silicon photonic platform, our theory may contribute to NH wave manipulation via metamaterial manufacture and laser technology. We also stress that our reflectionless property derives from making use of wave equation itself, which is distinct from unidirectional reflectionlessness from exceptional points in NH scattering system~\cite{LuoJ2018PT, Yuxuan2025GeneralizedPT, YanD2025Ultrasensitive}, and that our construction guarantees reflectionlessness only for the prescribed incident field at the design frequency, rather than works for a general class of incident waves.

\section{Theory}
The theory of inverting wave equation applicable to NH scenarios shall be outlined in Sec. II. The central result in our Letter in Eq.~\eqref{eq:central}. We start from the Helmholtz equation in two dimensions (2D) for simplicity~\cite{Vial2016class, Y2017Direct, King2018scattering-free}, which describes the wave transport for a linearly polarized electric field with wavelength $\lambda_0=2\pi/k_0$ in an isotropic material. And an incidence wave $E_{\rm in}(x, y)$ in air satisfying $\nabla^2E_{\rm in}+k_0^2E_{\rm in}=0$ will induce a total wave $E(x, y)$ passing through a medium with refractive index $n(x, y)$, which shall follows
\begin{equation}
\nabla^2E+k_0^2n^2(x, y)E=0. \label{eq:refname2}
\end{equation}
Here the total wave solution is made up of both incidence wave $E_{\rm in}(x, y)$ and scattering part $E_{\rm sc}(x, y)$~\footnote{Note that throughout this Letter wave flow is taken to impinge from left (the negative $x$ side) to progress toward the right end (positive $x$).}, 
\begin{equation}
E(x, y)=E_{\rm in}(x, y)+E_{\rm sc}(x, y). 
\label{eq:total}
\end{equation}
Using complex-valued wave function in~\eqref{eq:total}~\cite{Vial2016class}, the complex-valued medium $n(x, y)$ is naturally NH where the imaginary index stands for sink and source, obtained by inverting ~\eqref{eq:refname2}. 
\begin{equation}
n(x, y)=\sqrt{\frac{k_0^2E_{\rm in}-\nabla^2E_{\rm sc}}{k_0^2E}}.
\label{eq:central}
\end{equation}
Therefore as the central result of this Letter, \eqref{eq:central} gives the design recipe for media profile of reflectionless waves~\footnote{Here we take the root with the positive real part throughout this paper although the other could well work as the negative-index medium. And the scattered wave $E_{\rm sc}$ stays zero at the left side, and becomes nonzero in the middle and right region by the designed material, which clarifies why the left sides of our designed media always stay at unity index, asymptotically $n\vert_{x=-\infty}\sim 1$.}. 
%The wave solutions in \eqref{eq:total} are complex functions of frequency and space in polar form~\cite{Vial2016class}, thus naturally enabling the NH medium where the imaginary index stands for sink and source. 
This inverting method provides sufficient room to manipulate waves, which we shall illustrate via three types of examples in Sec.~\ref{Ex} respectively: controlling the amplitude, the phase, and the propagating direction. 

\section{Examples}\label{Ex}

\subsection{Amplitude Control}\label{Amp}

\begin{figure*}[htbp!]
	\centering
	%      \fbox{\includegraphics[width=0.95\textwidth]{Figure/fig1.png}}
	\fbox{\includegraphics[width=0.95\linewidth]{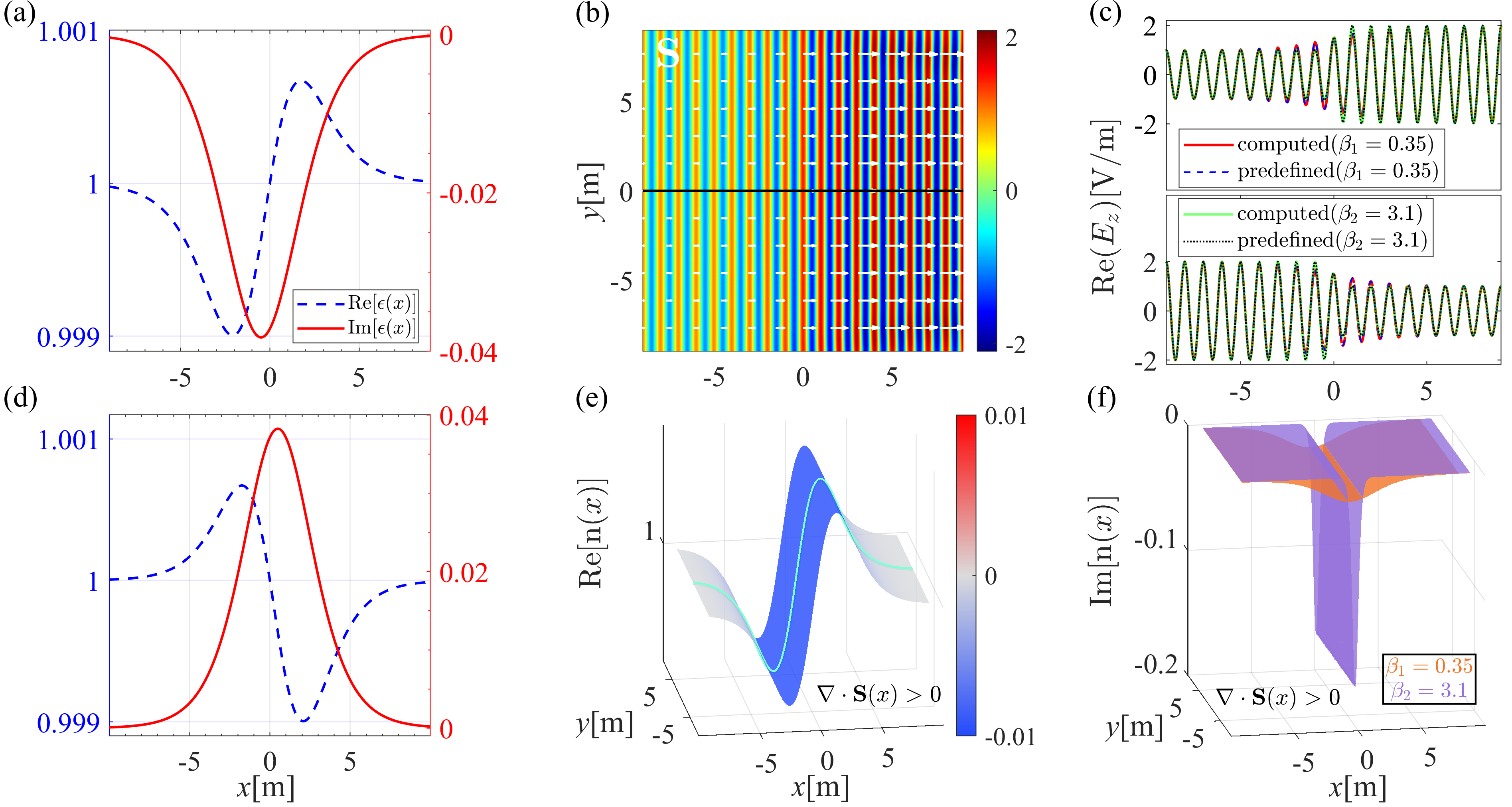}}
	\caption{Amplitude modulation of a plane wave by two NH media: (a) gain medium vs. (d) loss medium in permittivity profiles; (b) electric field in the gain medium with white arrows indicating the Poynting vector; (c) matched fields of computation and predefinition for both media, for varied $\beta$ values to tune; (e) schematic of wave propagation in the gain medium (a); (f) imaginary parts of refractive index under different $\beta$ values for the gain medium (a), which aligns with a positive divergence of Poynting vector $\nabla\cdot\mathbf{S}>0$ as a wave source. In panels (b-c) and all numerics below perfect matched layers (PML) are used to truncate the computational domain.  }
	\label{figs_fig1}
\end{figure*}

As a simplest example for modulating the amplitude, we imagine a plane wave with phase $\exp (ik_0 x)$ propagating through the design medium along $+x$ direction $\exp{(-i\omega t ) }$ as convection~\cite{Kresic2022Transforming}), and its amplitude variess smoothly from $A_1$ to $A_2$ for both ports of input and output while preserving its phase all along. Then the modulated plane wave shall carry a position-dependent amplitude $A(x)$: 
\begin{equation}
A(x)=A_1+(A_2-A_1)f\left(x\right), 
\label{eq:A}
\end{equation}
where $f\left(x\right) ={[1 + \tanh \left(\beta x\right)]}/{2}$ transits smoothly from $A_1$ to $A_2$. Then the wave solution in the medium is $E(x)=A(x) \exp(ik_0x)$. From \eqref{eq:total} the hence-scattered part should be 
\begin{equation}
E_{\rm sc}(x)=f\left(x\right)\left(A-A_{1}\right) \exp \left(ik_0x\right). 
\label{eq:refname7}
\end{equation}
Substituting $E_{\rm in}=A_1 \exp \left(ik_0x\right)$ and~\eqref{eq:refname7} into \eqref{eq:central} gives
\begin{widetext}
\begin{equation}
n(x)=\sqrt{1-\frac{(A-A_{1})f^{\prime\prime}(x)}{k_0^2[A_{1}+\\f\left(x\right)\left(A-A_{1}\right)]}-i\frac{2(A-A_{1})f^{\prime}(x)}{k_0[A_{1}+f\left(x\right)\left(A-A_{1}\right)]}}. 
\label{eq:refname8}
\end{equation}
\end{widetext} 
In this manner, the incidence wave of amplitude $A_1$ interacts within medium $n(x)$ progressively in \eqref{eq:refname8} and then results in another planar wave of amplitude $A_2$.

For a gain medium to increase $A_1=1\rm{V/m}$ to $A_2=2\rm{V/m}$, an index profile in Fig.~\ref{figs_fig1}(a) is obtained from~\eqref{eq:refname8}. 
%the following material parameter: 
%\begin{equation}
%  n^2=1-\frac{f^{\prime\prime}(x)}{k_0^2[1+f\left(x\right)]}-i\frac{2f^{\prime}(x)}{k_0[1+f\left(x\right)]}. 
%\label{eq:refname11}
%\end{equation}
And the wave field $E(x)$ walking through our design medium in Fig.~\ref{figs_fig1} (a) verifies our theory in \eqref{eq:refname8} in Fig.~\ref{figs_fig1} (b) and the upper Fig.~\ref{figs_fig1}(c) for $y=0$. One could also reverse the propagation direction to achieve a shrunk amplitude the other way round ($A_1=2{\rm V/m}, A_2=1\rm{V/m}$), as presented in the lower of Fig.~\ref{figs_fig1}(c). And this requires a loss medium in Fig.~\ref{figs_fig1}(d), whose real permittivity is reverse in space and whose imaginary permittivity is complex conjugate to the gain medium in Fig.~\ref{figs_fig1}(a).

\emph{Curved surface analogue}: the non-Hermitian (NH) medium can be understood as a curved, coloured surface in a nutshell as shown in Fig.~\ref{figs_fig1}(e), where its height depending on spatial dimensions represents the real part of the refractive index $\Re [n(x, y)]$~\cite{Y2013Motion}, and the gradient colour indicates its imaginary part $\Im [n(x, y)]$. Then the wave-flowing picture in Fig.~\ref{figs_fig1}(b) corresponds to the light trajectory in aquamarine on the curved surface in Fig. \ref{figs_fig1}(e), dictated by the complex refractive index $n(x, y)$. Note that the complex permittivity profiles in Figs.~\ref{figs_fig1}(a) and (d) both strikingly resemble the linear susceptibility by Lorentz dispersion of the atom~\cite{Boyd2025NL}. The trajectory follows the geodics on the surface governed by the real part, and the wave amplitude is controlled by the negative imaginary for energy gain during propagation in this case. Another tuning parameter for transition is parameter $\beta$ in $f(x)$, which directly controls the degree of amplitude variation for amplitude. This tuning effect is visible from  
the sharpened amplitude-varying, as shown in Fig. \ref{figs_fig1}(c) for tuning parameter $\beta$ from $0.35$ to $3.1$. And a close-up of imaginary refractive index in Fig.~\ref{figs_fig1}(f) reveals such a distinction: a steeper surface of the imaginary index in purple points to a more rapid amplification, while the gentler slope in orange indicates a slower one. For the loss and gain degree of freedom, we may speculate a relation $\Im n \approx - \nabla\cdot \mathbf{S}$~\cite{Kresic2022Transforming} so that a positive divergence of Poynting vector $\nabla\cdot\mathbf{S}>0$ in Fig.~\ref{figs_fig1}(f) represents a wave source with negative index $\Im n(x, y)<0$, and vice versa for a negative divergence $\nabla\cdot\mathbf{S}<0$ and $\Im n(x, y)>0$.

% positive divergence of Poynting vector $\nabla\cdot\mathbf{S}>0$ represents a wave source with negative index $Im n(x, y)<0$, and a negative divergence $\nabla\cdot\mathbf{S}<0$ does a wave sink with positive imaginary index $Im n(x, y)>0$.

%Hence the tuning parameter $\beta$ to for the rapidness of amplitude variation. 

\subsection{Phase conversion}\label{subsec:phase}
Other than amplitude control in Subsec.~\ref{Amp}, we now turn to manipulate the wave phase via three cases. 
For an incident planar wave, we are equipped to design its phase $S\left(x,y\right)$ to smoothly shift away from planar to otherwise.  So for a total wave $E = \exp(iS)$~\cite{Ossi2022topo_const} and from \eqref{eq:central} the designed medium is
\begin{equation}
n={k_0}^{-1}{\sqrt{(S_x^{\prime})^2+(S_y^{\prime})^2-i{(S_{xx}^{\prime\prime}+S_{yy}^{\prime\prime}})}}.
\label{eq:nS}
\end{equation}
We shall give three cases to illustrate the principle in \eqref{eq:nS}. The first medium modulates one wave number to another, and the second converts the linear phase to a quadratic front in a continuously-varying manner as it propagates, both of which works in one dimension. And the third works in 2D to convert phase from planar to cylindrical. In all cases in Subsec.~\ref{subsec:phase}, one designs the one-dimensional phase $S\left(x\right)$ to follow
\begin{equation}
S(x) = S_1+\left(S_2-S_1\right) f\left(x\right),
\label{eq:Sx}
\end{equation}
similar to \eqref{eq:A} where only $S_2$ is exemplified in each case. 

\begin{figure*}[ht]
	\centering
	\fbox{\includegraphics[width=0.95\linewidth]{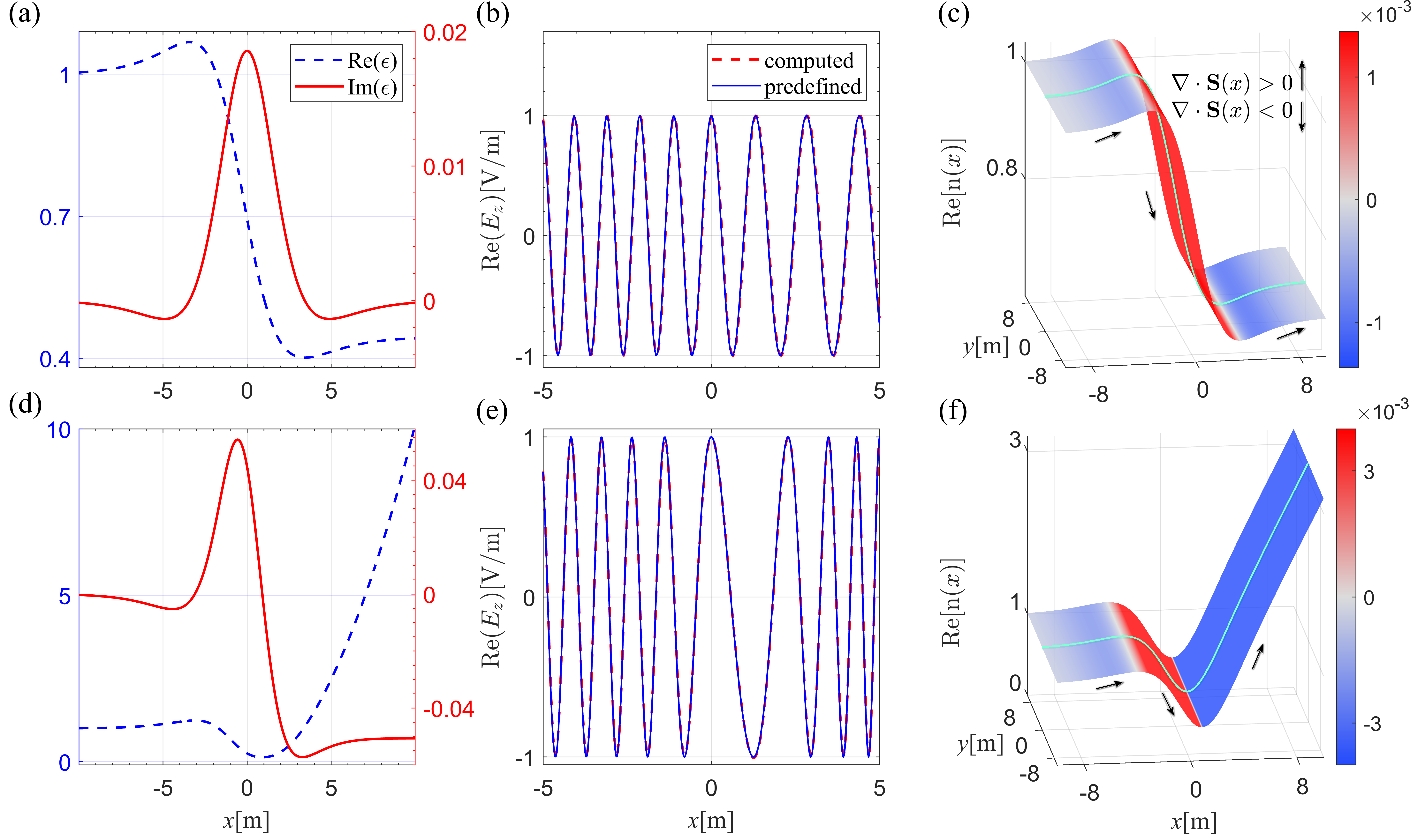}}
	\caption{Phase conversion from $k_0x$ to $k_1x$ (with $k_1={2k_0}/{3}$) via a NH medium: (a) distribution of the permittivity; (b) electric field along $x$ axis for comparison between computed and predefined fields; (c) curved surface analogue with coloured imaginary. Phase conversion from $k_0x$ to $k_0^2x^2$ similarly: (d) - (f). In panels (c) and (f), positive divergence of Poynting vector represents a wave source, and vice versa for sink similar to Fig.~\ref{figs_fig1}(f).  }
	\label{figs_fig2}
\end{figure*}

\emph{Modulated planar phase}: to design a medium that modifies the phase of an incident plane wave from $S_1=k_0x$ to $S_2=k_1x$, one yields the material profile illustrated in Fig.~\ref{figs_fig2}(a) by substituting \eqref{eq:Sx} into \eqref{eq:nS}. And the wave field thus-formed in Fig.~\ref{figs_fig2}(b) validates that the medium decreases the wave number from $k_0$ to $k_1$. The curved surface analogue for the conversion medium works for our defined incidence wave only, also illustrated in Fig.~\ref{figs_fig2}(c), where the ray propagates along the aquamarine line and the imaginary index is mainly lossy as a sink on the surface. 

\emph{Quadratic phase}: one may alter the phase from linear $k_0x$ to quadratic $k_0^2x^2$, i.e. using $S_2=k_0^2x^2$ to design a material shown in Fig.~\ref{figs_fig2}(d). The wave field in Fig.~\ref{figs_fig2}(e) then confirms such a phase conversion from linear to quadratic. And the surface picture in Fig. \ref{figs_fig2}(f) reveals that at the conversion region, a sink first and a source later are collaborated to achieve \emph{just} the quadratic phase as designed.

\emph{2D phase conversion-from planar to cylindrical}: we now convert the phase further in 2D, by designing to transform the planar phase into cylindrical as if through a lens~\cite{LuB2014Far-Zone, Isakov2016Directive}. The resultant phase $S(x,y)$ from \eqref{eq:Sx} is continuously varied to the exit phase $S_2$: 
\begin{equation}
S_2(x, y) =k_0\sqrt{(x+b)^2+y^2},
\label{eq13}
\end{equation}
as if emitted from the virtual source $(-b, 0)$. 

\begin{figure}[t]
	\centering
	\fbox{\includegraphics[width=0.98\linewidth]{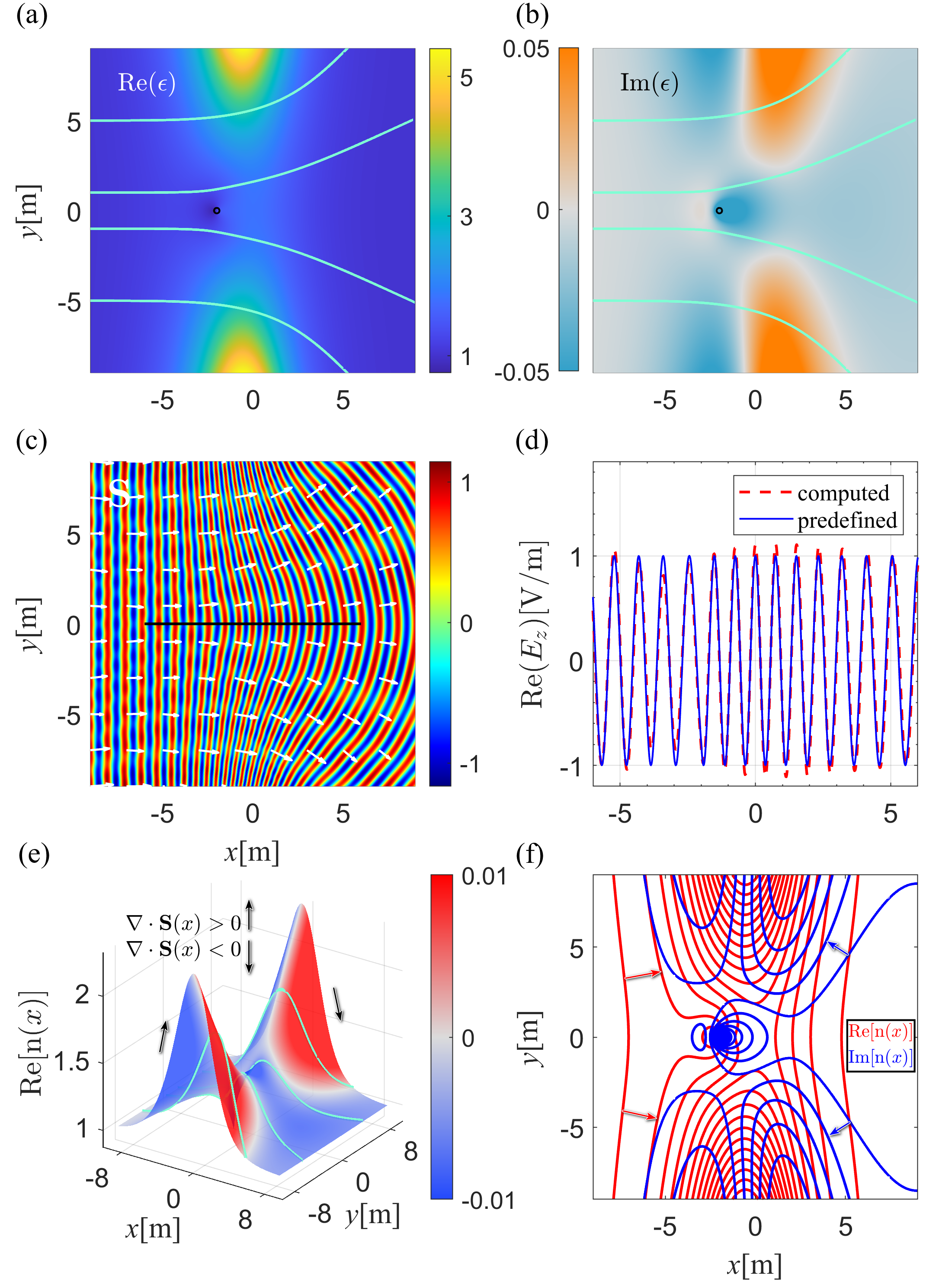}}
	\caption{Conversion of plane-wave phase to cylindrical-wave phase in a NH medium. (a, b) Distribution of the complex permittivity and note the virtual point $(-b, 0)$ marked by circles; (c) electric field distribution; (d) comparison between predefined and simulated electric fields along the black line in (c); (e) schematic of wave propagation through the designed medium; (f) contour plots showing the real and imaginary parts of the refractive index, both with an arrow indicating the increasing directions. Parameter: $b=2$. }
	\label{figs_fig3}
\end{figure}

By substituting \eqref{eq13} into \eqref{eq:nS} and \eqref{eq:Sx}, the real and the imaginary parts of the index profile are shown respectively in Figs. \ref{figs_fig3}(a) and (b), which is truncated by PML applied and would extends further along $y$-direction otherwise. As the plane wave walks through the designed medium, the electric field converts its phase from planar to cylindrical smoothly with unity amplitude, as shown in Figs.~\ref{figs_fig3}(c) and(d)~\footnote{Note that the calculated fields in Fig.~\ref{figs_fig3}(b) reveal slight discrepancy under (near $x=-3$) and over (near $x=1$) prediction, which we attribute to imperfect absorption of boundary-reflected waves by PML truncating the index profile which would extends further otherwise. As a convection, the PML is applied at the boundaries of the computational domain to effectively absorb outgoing waves and eliminate spurious reflections, thereby providing accurate numerics for wave propagation in the designed medium. }. We note that a passive medium~\cite{Vial2016class, Y2017Direct} cannot achieve similar waves which results in mismatched amplitude [see Sec. 1 and Fig. S1, Supplement 1]. This examples highlights the synergy power by distributed source and sink (imaginary permittivity) in Figs.~\ref{figs_fig3}(b) and (e). The NH curved surface in Fig.~\ref{figs_fig3}(e) hides a distinct feature for real ($\Re[n(x, y)]$, red line) and imaginary ($\Im[n(x, y)]$, blue line) parts of indices in Fig.~\ref{figs_fig3}(f): they are non-orthogonal and henceforth non-conformal to each other~\cite{Kresic2022Transforming}. Furthermore, the imaginary part of the refractive index exhibits a gain region near the source point [cf. Figs.~\ref{figs_fig3}(b) and (e)], in order to induce the cylindrical wave front from there. And later the loss region near $0<x<5$ works to damp down the wave field \emph{just} to maintain the unity amplitude predefined.

\section{Asymmetric transmission to shunt the wave direction}\label{sec:wave_direc}
With an envelope-back derivation, one finds that for a complex conjugate medium, wave fronts can be swapped to achieve reversal transmission~[see Sec. 2, Supplement 1, cf. Figs.~1(a-d)]. That being said, the very same medium will mold the inverted wave front from the other side different than the original one, and henceforth produces \emph{asymmetric} transmission itself~\cite{FanY2025Passive} similar to \cite{MaG2024Unidirectional}. Then we can only design a unidirectional transmitter~\cite{Jalas2013} to shunt the wave vector to a specific direction, i.e., in one direction to shape the wave as predicted but not in the other direction. It is noted that our asymmetric transmitter works in mode conversion and does maintains Lorentz reciprocity~\cite{Caloz2018EM_Nonreciprocity, Zangwill2013}, which is formally preserved in our linear, time-invariant, scalar media. Fortunately this scenario still boils down to the phase converter in Subsec.~\ref{subsec:phase}. Therefore, for a predefined wave phase of an obliquely exit vector defined by exit angle $\theta$, 
\begin{equation}
\mathbf{k_2}=k_0(\hat{x}\cos\theta+\hat{y}\sin\theta). 
\end{equation}
Thus one may design a NH medium under the periodic boundary condition along vertical $y$ axis, so as to mold such a shunt phase. However, the predicted shunting turns out to work only for discrete exit angle $\theta$, 
\begin{equation}
\theta= \arcsin \frac{2\pi m}{k_0\Lambda}. \label{theta}
\end{equation}
This exit angle derives from the grating equation for the transverse wave vector $k_{y} ={ 2\pi m} / {\Lambda}$, where $\Lambda$ is the length in vertical direction along one vertical period with $m$ an integer.

\begin{figure}[tbp]
	\centering
	\fbox{\includegraphics[width=0.98\linewidth]{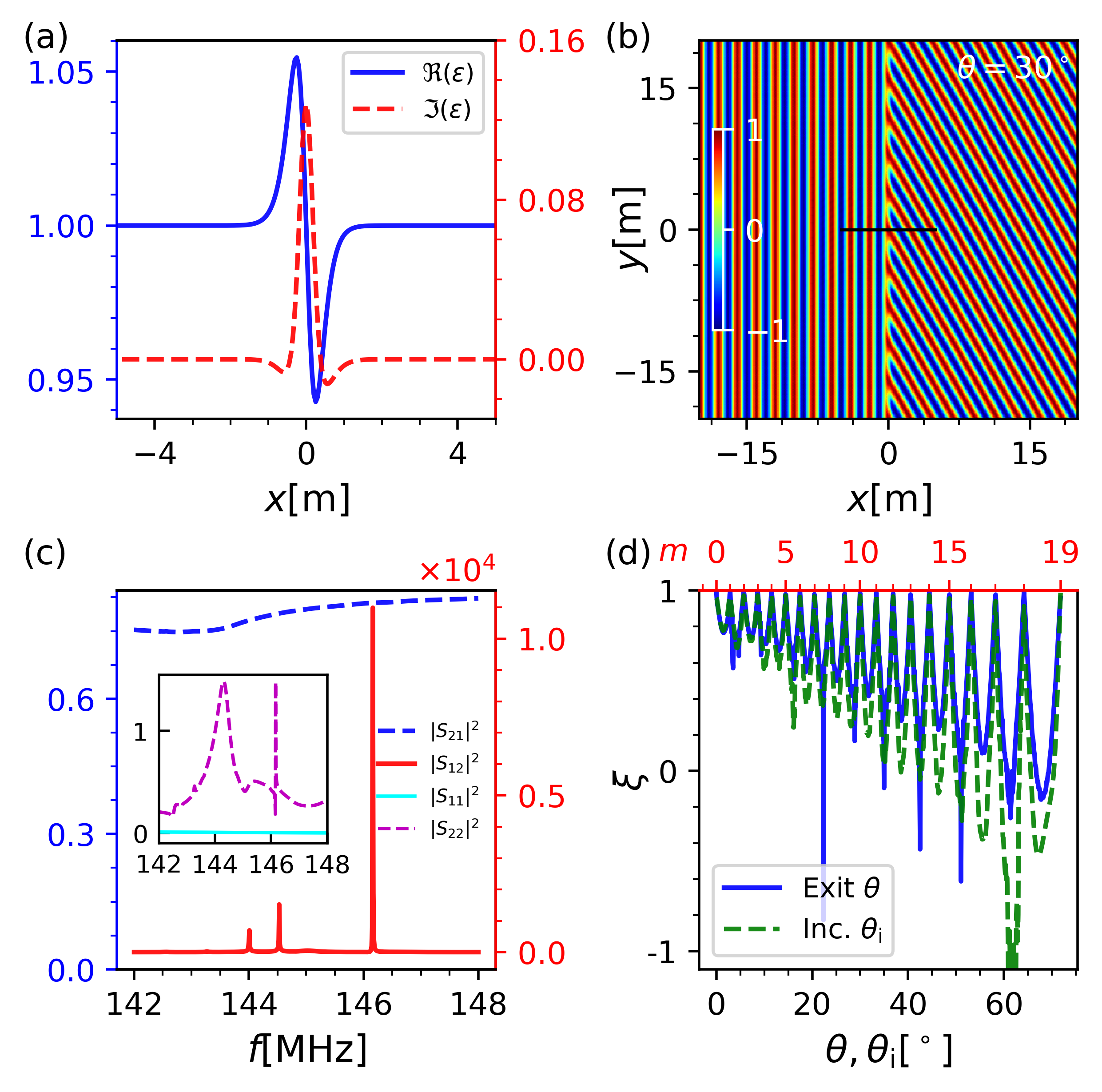}}
	\caption{Asymmetric transmission medium: (a) material parameter uniform in vertical direction; (b) electric field distribution when the incident plane wave is deflected from $\theta\vert m=0$ to exit at $\theta\vert m=10$; (c) four power ratios of reflection and transmission (inset) in scattering matrix $\mathbf{S}$; (d) relative accuracy of the calculated field $E_{\rm c}$ compared to the predefined $E_{\rm p}$ at different exit angles $\theta$ and incidence angles $\theta_{\mathrm{i}}$, both of which are defined in accordance to Eq.~\eqref{re_error}. } 
	\label{figs_fig4}
\end{figure}

An example of the angle $\theta\vert m=10$ is presented in Fig.~\ref{figs_fig4}. For a complex refractive index in Fig.~\ref{figs_fig4}(a) within a rather narrow region, the incident planar wave is converted towards the predicted direction $\theta(m=10)$, as shown in Fig.~\ref{figs_fig4}(b). To quantify the asymmetric transmission, we define the input/output channel as the power flowed through the left/right end, and its scattering matrix (2 by 2) is defined by
\begin{equation}
\begin{pmatrix}
b_\textbf{L}\\
b_\textbf{R}
\end{pmatrix}
=\begin{pmatrix}
S_{11} & S_{12}\\
S_{21} & S_{22}
\end{pmatrix}
\begin{pmatrix}
a_\textbf{L}\\
a_\textbf{R}
\end{pmatrix}, 
\end{equation}
in which $a_\text{L, R}$ stands for input power from the left/right end and $b_\text{L, R}$ for the output counterpart respectively. Here $a_\text{L}$ is the time-reversal mode of $a_\text{R}$ and $b_\text{L}$ is time-reversal of $b_\text{R}$~\cite{ZhangC2026Thesis}. So the pronounced asymmetric transmission due to mirror-symmetry broken profile~\cite{LuoJ2018PT} is confirmed by $S_{21}\lesssim 1$ and $S_{12}\approx 0$ except three eigenfrequency peaks~\cite{POOLE2016167} for a frequency range of $6~\mathrm{MHz}$ depicted in Fig.~\ref{figs_fig4}(c). Despite a mainly-lossy region near $x=0$, the unidirectional transmission works well for forward waves incident from the exit angle in \eqref{theta}, but severely suppresses back propagation~\cite{Garcia2016Nontensorial}. To access the shunting effects for other possible exit angles, the relative accuracy $\xi(\theta)$ of the calculated field from prediction, 
\begin{equation}
	\xi(\theta) = 1 - \frac{\displaystyle \left[\int_S (E_{\rm c} - E_{\rm p})(E_{\rm c}- E_{\rm p})^\ast \, dS\right]^{1/2}}{\displaystyle \left[\int_S E_{\rm p} E_{\rm p}^\ast \, dS\right]^{1/2}}, 
	\label{re_error} 
\end{equation}
peaks sharply at unity as shown in Fig.~\ref{figs_fig4}(d) [blue curves]. And intriguingly other discrete incidence angles $\theta_{\mathrm{i}}$ also work as if from a predefined incidence. This calculation validates the discrete angles given in \eqref{theta} [red dashed] due to the periodic boundary condition, and more instances are expected to be invented by our NH method. To evaluate the robustness of the designed functionality against angle deviation, more data are appended in Subsecs. 4B-4C, Supplement 1.

\section{Conclusion}

In a nutshell, our previous inverting method extends to the NH regime for designing isotropic media from known reflectionless waves, thus incorporating distributed source and sink. By using three types of examples we validate that this method not only avoids the index range below unity~\cite{Kresic2022Transforming}, but also facilitates precise manipulation of wave phases beyond conventional Hermitian paradigms~\cite{Y2017Direct, YanA2019Case} which dates back to construction of complex potentials in paraxial optics~\cite{Abdullaev2010Dissipative}. In principle, the achievable range of refractive index is put no explicit restraints on~\footnote{However, in metamaterial engineering we would need to consider dispersion issue for real index lower than 1. We would prefer to set the resultant profile more on the side of larger than one for real part, and not too much (magnitude less than 0.05 though it does not guarantee operation below lasing threshold) off zero for avoiding unnecessary nonlinear issue.}. And the proposed NH modulation could be experimented using spatially distributed absorber/lasing medium, which could boil down to discretised metamaterial~\cite{Vial2016class}. With the unique loss and gain effect~\cite{Komis2025imag_gauge}, this NH design toolkit may lead to further manipulation for light such as coherent perfect absorption~\cite{ChongY2010Coherent}, invisibility~\cite{Philbin2016All-freq}, and lasing~\cite{Bai2020NH_ph}, which shall all contribute to the design arsenal of inverse problem in nanophotonics~\cite{Piggott2015Inverse, Molesky2018Inverse}.

%\appendix % 

%\begin{backmatter}%\begin{acknowledgments} 

\section*{Acknowledgment}

Y. L. thanks Mao Zhu, Sun Zizhuang, Jin Luling, Chen Biao, Zeng Jinwei, Qiao Chang, and Han Song for their helpful discussions, and is especially grateful for the original idea by Simon Horsley from University of Exeter, UK. 

We are supported by National Natural Science Foundation of China (Grant Nos. 62571212, U25D8012), Science and Technology Department of Hubei Province (2024AFA038, 2022CFB553, 2022CFA012), Hubei International Cooperation Research Base of Quantum Optics and Devices (SH2416), Outstanding Young and Middle-aged Scientific and Technological Innovation Team of Colleges and Universities in Hubei Province (T2020001), Wuhan City Key R\&D program (2025050602030069), and 2023 supplemental grant for 1A0702E004: Modern Optics.

%\end{backmatter}

%\smallskip

%\noindent Here are examples of disclosures:

\emph{Disclosures} The authors declare no conflicts of interest. During the preparation of this work, the authors used ChatGPT for language revision. The authors reviewed and edited the output as needed and take full responsibility for the content of the published article.

\emph{Data availability} Data underlying the results presented in this paper are not publicly available at this time but may be obtained from the authors upon reasonable request.

\emph{Supplemental document}
See Supplement 1 for supporting content.

%\nocite{*}
\newpage

\bibliography{ZhangC_Refs3}

@book{Ulf2010Geometry,
   author = {Leonhardt, U. and Philbin, T.},
   title = {Geometry and light : the science of invisibility},
   publisher = {Dover Publications, Inc.},
   address = {Mineola, N.Y.},
   series = {Dover books on physics},
   pages = {278},
   ISBN = {9780486476933
0486476936},
   year = {2010},
   type = {Book}
}

@article{Horsley2015Spatial,
   author = {Horsley, S. A. R. and Artoni, M. and La Rocca, G. C.},
   title = {Spatial Kramers-Kronig relations and the reflection of waves},
   journal = {Nature Photonics},
   volume = {9},
   number = {7},
   pages = {436-439},
   ISSN = {1749-4885
1749-4893},
   DOI = {10.1038/nphoton.2015.106},
   year = {2015},
   type = {Journal Article}
}

@article{Ye2017reflectionless,
   author = {Ye, D. and Cao, C. and Zhou, T. and Huangfu, J. and Zheng, G. and Ran, L.},
   title = {Observation of reflectionless absorption due to spatial Kramers-Kronig profile},
   journal = {Nat Commun},
   volume = {8},
   number = {1},
   pages = {51},
   ISSN = {2041-1723 (Electronic)
2041-1723 (Linking)},
   DOI = {10.1038/s41467-017-00123-4},
   url = {https://www.ncbi.nlm.nih.gov/pubmed/28674391},
   year = {2017},
   type = {Journal Article}
}

@article{Molesky2018Inverse,
   author = {Molesky, S. and Lin, Z. and Piggott, A. Y. and Jin, W. and Vuckovic, J. and Rodriguez, A. W.},
   title = {Inverse design in nanophotonics},
   journal = {Nature Photonics},
   volume = {12},
   number = {11},
   pages = {659-670},
   ISSN = {1749-4885
1749-4893},
   DOI = {10.1038/s41566-018-0246-9},
   year = {2018},
   type = {Journal Article}
}

@article{Kresic2022Transforming,
  title = {Transforming Space with Non-Hermitian Dielectrics},
  author = {Kre\ifmmode \check{s}\else \v{s}\fi{}i\ifmmode \acute{c}\else \'{c}\fi{}, Ivor and Makris, Konstantinos G. and Leonhardt, Ulf and Rotter, Stefan},
  journal = {Phys. Rev. Lett.},
  volume = {128},
  issue = {18},
  pages = {183901},
  numpages = {6},
  year = {2022},
  month = {May},
  publisher = {American Physical Society},
  doi = {10.1103/PhysRevLett.128.183901}
}

@article{Y2017Direct,
doi = {10.1088/1367-2630/aa6c0c},
url = {https://dx.doi.org/10.1088/1367-2630/aa6c0c},
year = {2017},
month = {jul},
publisher = {IOP Publishing},
volume = {19},
number = {7},
pages = {073010},
author = {Liu, Y and Vial, B and Horsley, S A R and Philbin, T G and Hao, Y},
title = {Direct manipulation of wave amplitude and phase through inverse design of isotropic media},
journal = {New Journal of Physics}
}

@article{Vial2016class,
  title = {A class of invisible inhomogeneous media and the control of electromagnetic waves},
  author = {Vial, B. and Liu, Y. and Horsley, S. A. R. and Philbin, T. G. and Hao, Y.},
  journal = {Phys. Rev. B},
  volume = {94},
  issue = {24},
  pages = {245119},
  numpages = {6},
  year = {2016},
  month = {Dec},
  publisher = {American Physical Society},
  doi = {10.1103/PhysRevB.94.245119},
  url = {https://link.aps.org/doi/10.1103/PhysRevB.94.245119}
}

@article{Pendry2006Controlling,
author = {J. B. Pendry  and D. Schurig  and D. R. Smith },
title = {Controlling Electromagnetic Fields},
journal = {Science},
volume = {312},
number = {5781},
pages = {1780-1782},
year = {2006},
doi = {10.1126/science.1125907},
URL = {https://www.science.org/doi/abs/10.1126/science.1125907},
eprint = {https://www.science.org/doi/pdf/10.1126/science.1125907}}

@article{Kresic2021Confinement,
author = {Kre\ifmmode \check{s}\else \v{s}\fi{}i\ifmmode \acute{c}\else \'{c}\fi{}, I. and Makris, K. G. and Rotter, S.},
title = {Light Confinement by Local Index Tailoring in Inhomogeneous Dielectrics},
journal = {Laser \& Photonics Reviews},
volume = {15},
number = {9},
pages = {2100115},
doi = {https://doi.org/10.1002/lpor.202100115},
year = {2021}
}

@article{Makris2020Scattering-free,
author = {K. G. Makris and I. Kre\v{s}i\'{c} and A. Brandst\"{o}tter and S. Rotter},
journal = {Optica},
number = {6},
pages = {619--623},
publisher = {Optica Publishing Group},
title = {Scattering-free channels of invisibility across non-Hermitian media},
volume = {7},
month = {Jun},
year = {2020},
url = {https://opg.optica.org/optica/abstract.cfm?URI=optica-7-6-619},
doi = {10.1364/OPTICA.390788},
}

@article{El2018NH_phys,
   author = {El-Ganainy, R. and Makris, K. G. and Khajavikhan, M. and Musslimani, Z. H. and Rotter, S. and Christodoulides, D. N.},
   title = {Non-Hermitian physics and PT symmetry},
   journal = {Nature Physics},
   volume = {14},
   number = {1},
   pages = {11-19},
   ISSN = {1745-2473
1745-2481},
   DOI = {10.1038/nphys4323},
   year = {2018},
   type = {Journal Article}
}

@article{Isakov2016Directive,
   author = {Isakov, D. and Stevens, C. J. and Castles, F. and Grant, P. S.},
   title = {3D-Printed High Dielectric Contrast Gradient Index Flat Lens for a Directive Antenna with Reduced Dimensions},
   journal = {Advanced Materials Technologies},
   volume = {1},
   number = {6},
   pages = {1600072},
   ISSN = {2365-709X},
   DOI = {https://doi.org/10.1002/admt.201600072},
   url = {https://doi.org/10.1002/admt.201600072},
   year = {2016},
   type = {Journal Article}
}

@article{LuB2014Far-Zone,
   author = {Lu, B. and Jiang, Z. and Werner, D. H.},
   title = {Far-Zone Focusing Lenses Designed by Complex Coordinate Transformations},
   journal = {IEEE Antennas and Wireless Propagation Letters},
   volume = {13},
   pages = {1779-1782},
   ISSN = {1536-1225
1548-5757},
   DOI = {10.1109/lawp.2014.2356173},
   year = {2014},
   type = {Journal Article}
}

@article{Garcia2016Nontensorial,
   author = {Garcia-Meca, C. and Barcelo, C.},
   title = {Nontensorial Transformation Optics},
   journal = {Physical Review Applied},
   volume = {5},
   number = {6},
   ISSN = {2331-7019},
   DOI = {10.1103/PhysRevApplied.5.064008},
   year = {2016},
   type = {Journal Article}
}

@book{Boyd2025NL,
   author = {Boyd, Robert W.},
   title = {Nonliear Optics},
   publisher = {Beijing World Publishing Corporation},
   address = {Beijing},
   edition = {4th ed.},
   year = {2025},
   type = {Book}
}

@article{Jalas2013,
    author = {Jalas, D. and Petrov, Al. and Eich, M. and Freude, W. and Fan, S. and Yu, Z. and Baets, R. and Popovi{\'c}, M. and Melloni, A. and Joannopoulos, J. D. and Vanwolleghem, M. and Doerr, C. R. and Renner, H.},
    title = {What is - and what is not - an optical isolator},
    journal = {Nature Photonics},
    volume = {7},
    number = {8},
    pages = {579--582},
    year = {2013},
    month = {08},
    date = {2013/08/01},
    issn = {1749-4893},
    doi = {10.1038/nphoton.2013.185},
    url = {https://doi.org/10.1038/nphoton.2013.185}
}

@incollection{POOLE2016167,
title = {Chapter 6 - S-parameters},
editor = {Clive Poole and Izzat Darwazeh},
booktitle = {Microwave Active Circuit Analysis and Design},
publisher = {Academic Press},
address = {Oxford},
pages = {167-204},
year = {2016},
isbn = {978-0-12-407823-9},
doi = {https://doi.org/10.1016/B978-0-12-407823-9.00006-8},
url = {https://www.sciencedirect.com/science/article/pii/B9780124078239000068},
author = {Clive Poole and Izzat Darwazeh}
}

@article{King2018scattering-free,
   author = {King, C.  and Horsley, S.  and Philbin, T.},
   title = {Designing scattering-free isotropic index profiles using phase-amplitude equations},
   journal = {Phys. Rev. A},
   volume = {97},
   pages = {053818},
   url = {http://adsabs.harvard.edu/abs/2018arXiv180202236K},
   year = {2018},
   type = {Journal Article}
}

@article{YanA2019Case,
   author = {Yan, A. and Liu, Y. and Wang, W.},
   title = {Case study: A simple optical inverse problem from a geometrical optics point of view},
   journal = {Journal of Advanced Dielectrics},
   pages = {1950019},
   ISSN = {2010-135X},
   DOI = {10.1142/S2010135X1950019X},
   url = {https://doi.org/10.1142/S2010135X1950019X},
   year = {2019},
   type = {Journal Article}
}

@article{Piggott2015Inverse,
   author = {Piggott, A. Y. and Lu, J. and Lagoudakis, K. G. and Petykiewicz, J. and Babinec, T. M. and Vuckovic, J.},
   title = {Inverse design and demonstration of a compact and broadband on-chip wavelength demultiplexer},
   journal = {Nat. Photon.},
   volume = {9},
   number = {6},
   pages = {374-377},
   ISSN = {1749-4885},
   DOI = {10.1038/nphoton.2015.69
http://www.nature.com/nphoton/journal/v9/n6/abs/nphoton.2015.69.html#supplementary-information},
   url = {http://dx.doi.org/10.1038/nphoton.2015.69}, 
   year = {2015},
   type = {Journal Article}
}

@article{Y2013Motion,
   author = {Liu, Y. and Ang, L. K.},
   title = {Motion-induced radiation from electrons moving in Maxwell's fish-eye},
   journal = {Scientific Reports},
   volume = {3},
   number = {1},
   pages = {3065},
   ISSN = {2045-2322},
   DOI = {10.1038/srep03065},
   url = {https://doi.org/10.1038/srep03065},
   year = {2013},
   type = {Journal Article}
}

@article{Philbin2016All-freq,
   author = {Philbin, T. G.},
   title = {All-frequency reflectionlessness},
   journal = {Journal of Optics},
   volume = {18},
   number = {1},
   pages = {01LT01},
   ISSN = {2040-8978
2040-8986},
   DOI = {10.1088/2040-8978/18/1/01lt01},
   year = {2016},
   type = {Journal Article}
}

@article{Philbin2014Making,
   author = {Philbin, T. G.},
   title = {Making geometrical optics exact},
   journal = {Journal of Modern Optics},
   volume = {61},
   pages = {552},
   ISSN = {0950-0340
1362-3044},
   DOI = {10.1080/09500340.2014.899646, http://arxiv.org/abs/1402.2811},
   url = {http://arxiv.org/abs/1402.2811},
   year = {2014},
   type = {Journal Article}
}

@article{YuS2018Bohmian,
   author = {Yu, S. and Piao, X. and Park, N.},
   title = {Bohmian Photonics for Independent Control of the Phase and Amplitude of Waves},
   journal = {Phys. Rev. Lett.},
   volume = {120},
   number = {19},
   pages = {193902},
   ISSN = {1079-7114 (Electronic)
0031-9007 (Linking)},
   DOI = {10.1103/PhysRevLett.120.193902},
   url = {https://www.ncbi.nlm.nih.gov/pubmed/29799257},
   year = {2018},
   type = {Journal Article}
}

@article{Bai2020NH_ph, 
   author = {Bai, P. and Luo, J. and Chu, H. and Lu, W. and Lai, Y.},
   title = {Non-Hermitian photonics for coherent perfect absorption, invisibility, and lasing with different orbital angular momenta},
   journal = {Optics Letters},
   volume = {45},
   number = {24},
   pages = {6635-6638},
   DOI = {10.1364/OL.409690},
   url = {https://opg.optica.org/ol/abstract.cfm?URI=ol-45-24-6635},
   year = {2020},
   type = {Journal Article}
}

@article{Ossi2022topo_const,
   author = {Ossi, N. and Chandramouli, S. and Musslimani, Z. H. and Makris, K. G.},
   title = {Topological constant-intensity waves},
   journal = {Optics Letters},
   volume = {47},
   number = {4},
   pages = {1001-1004},
   DOI = {10.1364/OL.441942},
   url = {https://opg.optica.org/ol/abstract.cfm?URI=ol-47-4-1001},
   year = {2022},
   type = {Journal Article}
}

@article{ChongY2010Coherent,
   author = {Chong, Y. D. and Ge, L. and Cao, H. and Stone, A. D.},
   title = {Coherent perfect absorbers: time-reversed lasers},
   journal = {Phys. Rev. Lett.},
   volume = {105},
   number = {5},
   pages = {053901},
   ISSN = {1079-7114 (Electronic)
0031-9007 (Linking)},
   DOI = {10.1103/PhysRevLett.105.053901},
   url = {https://www.ncbi.nlm.nih.gov/pubmed/20867918},
   year = {2010},
   type = {Journal Article}
}

@article{FanY2025Passive,
   author = {Fan, Y. and Zhang, S. and Zong, M. and Liu, Y. and Lv, J. and Xu, Z.},
   title = {Passive non-reciprocal metasurfaces based on independently tunable nonlinear dual bound states in the continuum},
   journal = {Optics Letters},
   volume = {50},
   number = {8},
   pages = {2651-2654},
   DOI = {10.1364/OL.553156},
   url = {https://opg.optica.org/ol/abstract.cfm?URI=ol-50-8-2651},
   year = {2025},
   type = {Journal Article}
}

@article{Biswas2024Reflectionless,
   author = {Biswas, S. S. and Remesh, G. and Achanta, V. G.l and Banerjee, A. and Ghosh, N. and Gupta, S. D.},
   title = {Reflectionless propagation of beams through a stratified medium},
   journal = {Optics Communications},
   volume = {569},
   pages = {130766},
   ISSN = {0030-4018},
   DOI = {https://doi.org/10.1016/j.optcom.2024.130766},
   url = {https://www.sciencedirect.com/science/article/pii/S0030401824005030},
   year = {2024},
   type = {Journal Article}
}

@article{Steinfurth2022ph_constant-int,
   author = {Steinfurth, A. and Kresic, I. and Weidemann, S. and Kremer, M. and Makris, K. G. and Heinrich, M. and Rotter, S. and Szameit, A.},
   title = {Observation of photonic constant-intensity waves and induced transparency in tailored non-Hermitian lattices},
   journal = {Science Advances},
   volume = {8},
   number = {21},
   pages = {eabl7412},
   DOI = {10.1126/sciadv.abl7412},
   url = {https://doi.org/10.1126/sciadv.abl7412},
   year = {2022},
   type = {Journal Article}
}

@article{Komis2025imag_gauge,
  title={Effect of imaginary gauge on wave transport in driven-dissipative systems},
  author={I. Komis and K. G. Makris and K. Busch and R. El-Ganainy},
  journal={Physical Review A},
  year={2025},
  url={https://api.semanticscholar.org/CorpusID:276929164}
}

@article{Abdullaev2010Dissipative,
   author = {Abdullaev, F. Kh. and Konotop, V. V. and Salerno, M. and Yulin, A. V.},
   title = {Dissipative periodic waves, solitons, and breathers of the nonlinear Schrodinger equation with complex potentials},
   journal = {Phys Rev E},
   volume = {82},
   number = {5 Pt 2},
   pages = {056606},
   ISSN = {1550-2376 (Electronic)
1539-3755 (Linking)},
   DOI = {10.1103/PhysRevE.82.056606},
   url = {https://www.ncbi.nlm.nih.gov/pubmed/21230612},
   year = {2010},
   type = {Journal Article}
}

@article{MaG2024Unidirectional,
   author = {Ma, G. and Shen, C.-Y. and Li, J. and Huang, L. and Isıl, C. and Ardic, F. O. and Yang, X. and Li, Y. and Wang, Y. and Rahman, Md S. S. and Ozcan, A.},
   title = {Unidirectional imaging with partially coherent light},
   journal = {Advanced Photonics Nexus},
   volume = {3},
   number = {06},
   ISSN = {2791-1519},
   DOI = {10.1117/1.Apn.3.6.066008},
   year = {2024},
   type = {Journal Article}
}

@article{LuoJ2018PT,
   author = {Luo, J. and Li, J. and Lai, Y.},
   title = {Electromagnetic Impurity-Immunity Induced by Parity-Time Symmetry},
   journal = {Physical Review X},
   volume = {8},
   number = {3},
   ISSN = {2160-3308},
   DOI = {10.1103/PhysRevX.8.031035},
   year = {2018},
   type = {Journal Article}
}

@article{Yuxuan2025GeneralizedPT,
   author = {Liu, Y. and Mei, R. and Zhang, Y. and Luo, J.},
   title = {Generalized Parity-Time Symmetric Metasurfaces},
   journal = {ACS Photonics},
   volume = {12},
   number = {12},
   pages = {6690-6697},
   ISSN = {2330-4022},
   DOI = {10.1021/acsphotonics.5c01730},
   url = {https://doi.org/10.1021/acsphotonics.5c01730},
   year = {2025},
   type = {Journal Article}
}

@article{YanD2025Ultrasensitive,
   author = {Yan, D. and Shalin, A. S. and Wang, Y. and Lai, Y. and Xu, Y. and Hang, Z. H. and Cao, F. and Gao, L. and Luo, J.},
   title = {Ultrasensitive Higher-Order Exceptional Points via Non-Hermitian Zero-Index Materials},
   journal = {Phys Rev Lett},
   volume = {134},
   number = {24},
   pages = {243802},
   ISSN = {1079-7114 (Electronic)
0031-9007 (Linking)},
   DOI = {10.1103/18gg-gvzc},
   url = {https://www.ncbi.nlm.nih.gov/pubmed/40742949},
   year = {2025},
   type = {Journal Article}
}

@article{Caloz2018EM_Nonreciprocity,
   author = {Caloz, C. and Alu, A. and Tretyakov, S. and Sounas, D. and Achouri, K. and Deck-Leger, Z.},
   title = {Electromagnetic Nonreciprocity},
   journal = {Physical Review Applied},
   volume = {10},
   number = {4},
   pages = {047001},
   DOI = {10.1103/PhysRevApplied.10.047001},
   url = {https://link.aps.org/doi/10.1103/PhysRevApplied.10.047001},
   year = {2018},
   type = {Journal Article}
}

@book{Zangwill2013,
  author    = {Zangwill, Andrew},
  title     = {Modern Electrodynamics},
  publisher = {Cambridge University Press},
  address   = {Cambridge},
  year      = {2013},
  pages     = {xxi, 977},
  isbn      = {9780521896979}
}

@MASTERSTHESIS{ZhangC2026Thesis,
  author = {C. Zhang},
  title = {{Controlling Electromagnetic Waves by Inverse Design of Non-Hermitian Media}},
  school = {Hubei University},
  year = {2026},
  address = {Wuhan},
}
\end{document}